\documentstyle[12pt,graphicx,pstricks,braket,amsmath,amssymb,enumitem,
                   float, epstopdf,pdfpages,cite,tabu,hyperref,subfigure, float,tikz]{article}
                    \usetikzlibrary{arrows,matrix,positioning}
\begin{document}
	\def\NPB#1#2#3{{\it Nucl.\ Phys.}\/ {\bf B#1} (#2) #3}

\def\AEF{A.E. Faraggi}

\begin{titlepage}
\samepage{
\setcounter{page}{1}
\rightline{}
\vspace{1.5cm}

\begin{center}
 {\Large \bf Comment on \cite{Marinho:2019zny}}
\end{center}

\begin{center}
%\vspace{1.cm}

%\vfill 
{\large
Johar M. Ashfaque$^\spadesuit$
}\\
\vspace{1cm}
$^\spadesuit${\it  Max Planck Institute for Software Systems,
            Campus E1 5,\\ 66123 Saarbr\"ucken, Germany\\}
\end{center}

\begin{abstract}
In this note, we show that the $\mu$-oscillator does not lead to the Fibonacci sequence as claimed in \cite{Marinho:2019zny} since $[n]^{\mu}=n$ in the limit $\mu \rightarrow 0$. Thus we obtain the sequence $[0]^{\mu}=0, 1, 2,...$. We only obtain the Fibonacci sequence when the $q$-deformation is associated to the $\mu$-deformation via the basic number 
$$\lim_{\mu\rightarrow 0} [n]^{\mu}_{q_1, q_2} = \frac{q_1^{2n}-q_2^{2n}}{q_1^2-q_2^2}.$$
\end{abstract}
\smallskip}
\end{titlepage}

\section{Comment}
In this note, we show that the $\mu$-oscillator does not necessarily lead to the Fibonacci sequence as claimed in \cite{Marinho:2019zny} since $[n]^{\mu}=n$ in the limit $\mu \rightarrow 0$.

The $q$-oscillator Fibonacci basic number necessarily satisfies the Fibonacci relation since 
$$[0]=0$$
$$[1]=1$$
$$[2]=\alpha$$
$$[3]=\alpha^2+\beta$$
$$.$$
$$.$$
$$.$$
and setting $\alpha=\beta=1$ yields the Fibonacci sequence.

However, when considering the $\mu$-oscillator 
$$[n]^{\mu} = \frac{n}{1+n\mu}$$
giving
$$[0]^{\mu}=0$$
and 
$$[1]^{\mu} = \frac{1}{1+\mu}\neq 1.$$
It is only in the limit $\mu \rightarrow 0$ that 
$$\lim_{\mu \rightarrow 0}[1]^{\mu} = 1.$$
Now, since  
$$\lim_{\mu \rightarrow 0}[n]^{\mu} =n$$
we have
$$\lim_{\mu \rightarrow 0}[2]=2$$
$$.$$
$$.$$
$$.$$

\end{document}